\documentclass[aps,pra,twocolumn,showpacs,preprintnumbers,amsmath,amsfonts,amssymb,footinbib,floatfix]{revtex4-2}
\pdfoutput=1
\usepackage{graphicx,graphics,epsfig,subfigure,times,bm,bbm,amssymb,amsmath,amsfonts,amsthm,mathrsfs,MnSymbol}
\usepackage{gensymb}
\usepackage[matrix,frame,arrow]{xypic}
\usepackage[pdfstartview=FitH]{hyperref}
\usepackage{subfigure}
\usepackage{comment}
\hypersetup{
    colorlinks=true,       % false: boxed links; true: colored links
    linkcolor=red,          % color of internal links
   citecolor=magenta,        % color of links to bibliography
    filecolor=magenta,      % color of file links
    urlcolor=cyan,           % color of external links
    runcolor=cyan
}
\usepackage[pdftex]{color}
\usepackage{braket}%Dirac Notation in QM
\usepackage{enumerate}
\usepackage[normalem]{ulem}
\usepackage[usenames,dvipsnames]{xcolor}
\usepackage{multirow}
\usepackage{mathtools}
\usepackage[T1]{fontenc}
\usepackage[utf8]{inputenc}

\definecolor{orange}{rgb}{1,0.5,0}

\newcommand{\bes} {\begin{subequations}}
\newcommand{\ees} {\end{subequations}}
\newcommand{\bea} {\begin{eqnarray}}
\newcommand{\eea} {\end{eqnarray}}

\definecolor{gold}{rgb}{0.85,.66,0}

\newcommand{\beq}{\begin{equation}}

\newcommand{\eeq}{\end{equation}}

\newcommand{\ignore}[1]{}

\def\bf{\mathbf{f}}

\def\s{\sigma}

\def\>{\rangle}
\def\<{\langle}

\def\Pr{\mathrm{Pr}}

\def\s0{I}

\newcommand{\ig}[1]{}

\begin{document}
%\begin{comment}
\title{Quantifying the Functional Size of Large-Scale Quantum Networks}
\author{Md Sohel Mondal$^1$, Vladimir S. Malinovsky$^2$, Siddhartha Santra$^1$\\\textit{$^1$ Department of Physics and CoE-QUICST, Indian Institute of Technology Bombay, Mumbai, Maharashtra 400076, India}\\ \textit{$^2$ DEVCOM Army Research Laboratory, 2800 Powder Mill Road, Adelphi, MD 20783, USA}}
\begin{abstract}
Large-scale quantum networks enable distributed quantum tasks by sharing entanglement between distant nodes. Since entanglement distribution in such networks relies on stochastic local operations and classical communication (SLOCC), the resulting end-to-end entanglement degrades with path length and exhibits statistical variations, limiting the usable range while also making the quantification of its functional size challenging. Here, we develop a statistical framework to quantify a network's functional size by analyzing the statistics of end-to-end entanglement connections established through arbitrary entanglement-distribution protocols. This framework yields a global network metric in the form of the graph radius of viable regions up to which the distributed entanglement meets specified task thresholds. In addition to identifying absolute viable regions, within which
a task is guaranteed to satisfy the prescribed thresholds, the framework identifies probabilistic viable regions, within which the task succeeds with a specified confidence. Further, it yields the viable node fraction, defined as the expected fraction of network nodes accessible from a given source node for the task, and reveals that network topology strongly influences the number of nodes contained within a network’s viable region, even when the viable radii are identical. Applying the framework to quantum key distribution on random networks
reveals sharp threshold behavior in the viable node fraction showing that small improvements in elementary-link
quality can produce a rapid increase in the viable node fraction of the network. Viable regions in a quantum network obtained using our statistical framework therefore provide a task-oriented characterization of quantum-
network functionality allowing comparison of  entanglement-distribution strategies
and network designs.
\end{abstract}
\maketitle
%\end{comment}

%%%%%%%%%%%%%%%%%%%%%%%%%%%%%%%%
%A quantum network (QN) \cite{unite_q_internet, Wehnereaam9288,towards_qn} utilises long-range entanglement as the essential resource for classically impossible tasks ranging from unconditionally secure communications to distributed quantum computation \cite{Wehnereaam9288, qnet_clocks, qkd1120, dist_q_comp_main, dist_q_comp_gomez}. Recent work has revealed interesting graph- and information-theoretic properties of such networks \cite{QN_stat_prop, QN_sat_brito, QN_cap_trans, robustness_noisy-networks, conc_perc_th, path_percolation}. However, the crucial practical issue of the effective size of a quantum network useful for a specific quantum task remains insufficiently explored due to the stochastic nature of the entanglement manipulation processes, viz. generation, swapping and purification, necessary for quantum networking \cite{azuma2023quantum}.

A quantum network (QN) \cite{unite_q_internet, Wehnereaam9288,towards_qn} provides long-range entanglement as an essential resource for powerful information processing tasks ranging from unconditionally secure communications to distributed quantum computation \cite{Wehnereaam9288, qnet_clocks, qkd1120, dist_q_comp_main}. This resource is obtained by combining short-ranged entangled states generated over the edges of the network
through probabilistic local operations and classical communication
(SLOCC) at the network nodes. The entanglement of the resulting end-to-end state degrades with the
number of elementary links (edges) involved in a long-range path while also showing statistical variations
due to the distribution of edge-parameters and the stochasticity of the entanglement manipulation operations \cite{swapping_sohel}. 

Further, since the ultimate purpose of a quantum network is to execute a quantum task, entanglement connections in the network must be assessed against the specific requirements that the task imposes on multiple parameters of the distributed entanglement \cite{qcm_sohel, mpep_mondal}. For example, a QKD protocol may require both a minimum fidelity and probability of successfully establishing the corresponding connection \cite{Wehnereaam9288}.

Together, these characteristics and requirements of long-range distributed entanglement in a quantum network present challenges in the quantification of its functional size, intuitively, the radius of the network region around a source node up to which the entanglement connections satisfy the requirements of a given quantum information processing (QIP) task. The functional size can serve as a useful network-level characterization of a quantum network accounting for not just graph connectivity of network regions but also its elementary-link properties, stochasticity of entanglement manipulation processes, and the requirements of the quantum task being performed \cite{QN_stat_prop, QN_cap_trans, robustness_noisy-networks, conc_perc_th}.

 %The functional size of a quantum network is therefore determined by a combination of topology, elementary-link properties, stochasticity of the entanglement manipulation processes, and the requirements of the quantum task being performed.

%The challenge in quantifying the functional size of a quantum network stems from the statistically rich behaviour of the end-to-end entanglement distribution. Although the average end-to-end entanglement generally decreases with source-destination separation \cite{swapping_sohel}, entanglement connections between node pairs at a fixed graph distance can exhibit substantial statistical variations due to the stochastic nature of the network edges and the entanglement-distribution process. Consequently, the probability that a source node can establish an entanglement connection satisfying the requirements of a given quantum information processing (QIP) task depends on its distance from the destination. Assessing how far one can go from a source node while maintaining a desired confidence of performing a QIP task therefore requires a statistical characterization of the end-to-end entanglement.

%which is defined as the graph distance one can go from a source node while maintaining a desired confidence of performing a QIP task.

In this letter, we propose a statistical framework to assess the functional size of a quantum network in terms of the radius of viable regions for QIP tasks, defined as network regions around an arbitrary source node within which the entanglement connections satisfy the requirements of the task with a desired confidence. An absolute viable region (AVR) represents the region where the task requirements are guaranteed to be satisfied, while a probabilistic viable region (PVR) extends this region to connections satisfying the requirements with a specified confidence. Thus, the AVR and PVR characterize the extent of a network, measured in terms of graph distance from arbitrary network nodes, over which a given QIP task can be performed at different confidence levels.

%family defined by the distribution of network edge-parameters on graphs of a given topology. 
The framework uses the probability distribution functions (PDFs) of the edge-parameters together with an entanglement-distribution protocol to obtain the PDF of the resulting end-to-end path-parameters connecting a source-destination pair. The path-length-dependent PDFs of the path parameters determine the probabilities that entanglement connections over varying length-ranges in the network satisfy the parameter threshold requirements of a given QIP task. Moreover, it enables estimation of the viable node fraction (VNF), which is the fraction of network nodes with which a given source node can perform the QIP task. Complementary to the AVR and PVR, which characterize viability in terms of graph distance, the VNF quantifies the fraction of the network that is usable for the task. Interestingly, the VNF can exhibit a sharp transition with respect to the edge-parameter quality, separating regions of low and high network functionality. 

%This statistical method characterizes the functional size of a network in terms of viable regions, defined as network regions around a source node within which the entanglement connections satisfy the requirements of a given QIP task. An absolute viable region (AVR) represents the region where the task requirements are guaranteed to be satisfied, while a probabilistic viable region (PVR) extends this region to connections satisfying the requirements with a specified confidence. Thus, the AVR and PVR characterize the spatial extent of a network over which a given QIP task can be performed at different confidence levels.

%We further characterize the functionality of a network through the viable node fraction (VNF), which gives the fraction of network nodes with which a given source node can perform the QIP task. Unlike the AVR and PVR, which characterize viability in terms of graph distance, the VNF quantifies the fraction of the network that is usable for the task. Interestingly, the VNF can exhibit a sharp transition in the space of edge-parameters, separating regions of low and high network functionality. %This indicates that a small improvement in the edge-parameters can lead to a large increase in the fraction of the network viable for a QIP task.

 These results describe an alternative, task-oriented framework to characterize the functional size of quantum networks, complementing existing studies based on average metrics \cite{QN_stat_prop} or entanglement percolation \cite{conc_perc_th}. While entanglement percolation identifies the critical edge thresholds required to establish macroscopic connectivity across an infinite network, practical quantum tasks require entanglement connections between specific node pairs that simultaneously satisfy multi-parameter thresholds. Furthermore, average connection metrics overlook the statistical variations in end-to-end states resulting from stochastic quantum operations. By evaluating the full PDFs of the path-parameters, our approach accounts for this stochasticity and quantifies the operational reach and functional size of the network through the AVR, PVR, and VNF.

We remark that statistical characterisation of network paths has already been shown to be useful for studying multiple aspects of quantum networking such as multi-path entanglement purification \cite{mpep_mondal}, optimally connecting quantum sub-networks \cite{subnet_shashank}, and assessing the quantum connectivity in quantum networks \cite{qcm_sohel} by some of the present authors. The results here develop this idea to answer the practically important question of determining the functional size of the imminent large-scale quantum networks. 
 
{\it Quantum network model}: We consider a QN defined on an underlying graph $G(V,E)$ where $V$ is the set of network nodes and the edges $e\in E$ represent entangled links in terms of two-qubit states $\rho_e$. The physical attributes of the network edges are described by a tuple of random-valued physical edge-parameters, $\boldsymbol{\mu}=(\mu_1,\dots,\mu_n)$, with $n$-the number of parameters that specify the distributed quantum state along the edges, such as the concurrence, probability of edge-entanglement, link latency etc. Without loss of generality, we choose the parameterization such that larger values of all path-parameters correspond to improved performance for the QIP task. In realistic scenario the microscopic details about the edge-parameters might be unknown, rather they follow certain distributions. In general, the $n$ different edge-parameters can be dependent random variables with a corresponding joint probability distribution function (PDF) $q_{\mu_1,\dots,\mu_n}^{(1)}$. A family of QNs, $\tilde{Q}$, is then fully specified given the tuple,
\begin{align}
\tilde{Q}:=\{G(V,E), q_{\mu_1,\dots,\mu_n}^{(1)}\},
\label{graph_dist}
\end{align}
along with an entanglement distribution protocol to establish an entangled link between any two nodes in the network.

An entanglement distribution protocol is required to establish an end-to-end entangled link between any two nodes $S$ and $D$. In general, the protocol can involve three probabilistic LOCC network operations: entanglement generation (EG), entanglement swapping (ES), and entanglement purification (EP). EG first establishes short-range entangled states, $\rho_e$, over the network edges \cite{ent_generation}. In general, EG may involve complex repeater protocols \cite{repgens,repeater_liang_2009,repeater_peter_2012,repeater_elkouss,repeater_peter_2022} that eventually generate entanglement between neighboring nodes. Rather than modeling these protocols in detail, we characterize their outcomes solely through the edge-parameters of the final entangled states they produce. The statistical features of these underlying protocols are therefore captured by the PDFs of the edge-parameters across the network. ES extends the range of entanglement by entangling network nodes connected through a path \cite{entswap_conc,swapping_sohel}. EP can further enhance the quality of entangled states by utilizing multiple entangled states in parallel. An entanglement distribution protocol can therefore utilize one or more end-to-end paths between $S$ and $D$ to obtain entangled states characterized by their respective \textit{path-parameters}, through a sequence of the three network operations. In this letter, we consider protocols based on single paths while our statistical framework presented below is applicable to any general entanglement distribution protocol.

{\it Characterizing entanglement connections in the network}: The entanglement connection between the node pair {\it S-D} via an entanglement distribution protocol, $\mathcal{E}$, in such a QN is fully described by an $n$-tuple, $\boldsymbol{\mu}_{SD}=(\mu_{1,SD},\mu_{2,SD},...,\mu_{n,SD})$, of random-valued path-parameters, $\mu_{i,SD}$. These path-parameters are determined by the edge-parameters along the path used for establishing the end-to-end entangled state. Therefore, for an arbitrary $S-D$ pair, they can be obtained as,
\begin{align}
    \mu_{i,SD}=\mathcal{E}_{i}\left(\{\boldsymbol{\mu}_{e}\}_{e\in \mathcal{P}_{SD}}\right),\qquad i\in\{1,\dots,n\},
\end{align}
where $\mathcal{P}_{SD}$ denotes the path used in the protocol and $\mathcal{E}_{i}$ denotes the protocol-dependent map from the complete set of edge-parameters along the path to the $i$th path-parameter.

In complex quantum networks, arbitrary pairs of nodes are in general characterized by their graph distance. For a given graph distance $l$, we consider the ensemble of node pairs $\mathcal{K}_l=\{(S,D)\mid d_G(S,D)=l\}$, where $d_G(S,D)$ denotes the graph distance between $S$ and $D$, i.e., the length of the shortest graph path (SGP) connecting them. The path-parameters $\mu_{i,SD}$ can differ among node pairs in $\mathcal{K}_l$ due to the inhomogeneity of the edge-parameters and the choice of the path used by the entanglement distribution protocol. Thus, the path-parameters associated with the node pairs in $\mathcal{K}_l$ can be characterized by a joint PDF, which we denote by $q_{\mu_1,\dots,\mu_n}^{(l)}$, and we denote the path-parameters of an arbitrary $(S,D)\in\mathcal{K}_l$ as $\mu_i^{(l)}\equiv \mu_{i,SD}$. Here, the superscript $l$ serves only to identify the ensemble of node pairs separated by graph distance $l$; it does not imply that the connection parameters are obtained from a path of length $l$, as the protocol may not utilize the SGP for entanglement distribution. For $l=1$, the connection parameters reduce to the corresponding edge-parameters, and $q_{\mu_1,\dots,\mu_n}^{(1)}$ is therefore the joint PDF of the edge-parameters considered in the preceding sections.
\begin{comment}
\begin{figure*}[t]
    \centering
    \includegraphics[width=\linewidth]{PRL submission/pdf_cp.pdf}
    \caption{(color online) Schematic illustration of the AVR and PVR around a source node (magenta) in a complex quantum network. The AVR and PVR, shown as green and yellow shaded regions, respectively, have radii $\tilde{r}_*=1$ and $r_*=2$ for a given QIP task.}
    \label{fig:pdf_cp}
\end{figure*}
\end{comment}

{\it Task-specific viability in a network}: In a quantum network, an arbitrary pair of nodes desires to perform a QIP task using the created end-to-end entanglement connection between them. Let us consider an arbitrary QIP task with required multiparameter threshold $(\mu_1^*,\dots,\mu_n^*)$. Such a QIP task is viable between a pair of network nodes if the path-parameters of the entanglement connection between the node pair, satisfies the threshold requirement for the task. In a network arbitrary node pairs separated by graph distance $l$ satisfies the QIP threshold statistically, with an associated task-specific \emph{viable probability} in the network. The viable probability is defined as the probability that the path-parameters exceed the multiparameter threshold. That is,
\begin{align}
   \eta_l:=\text{Pr}\left[\mu_1^{(l)}\geq\mu_1^*,\dots,\mu_n^{(l)}\geq\mu_n^*\right].
\end{align}
The quantity $\eta_l$ is different at different graph distances and can be obtained from the joint PDF of the path-parameters as follows,
\begin{align}
    \eta_l=\int^{\mu_{1,\text{max}}^{(l)}}_{\mu_1^*}\dots \int^{\mu_{n,\text{max}}^{(l)}}_{\mu_n^*} q_{\mu_1,\dots,\mu_n}^{(l)}~d\mu_1\dots d\mu_n,
    \label{viab_prob}
\end{align}
where $\mu_{i,\text{max}}^{(l)}$ denotes the maximum possible value of the $i^\text{th}$ path-parameter.
%The viable probabilities given in Eq.~(\ref{viab_prob}) corresponding to concurrence and probability for the thresholds $(c_*,p_*)=(0.8,0.016)$ and uniform distributions of the edge-concurrence $[U(0.9,1)]$ and edge-probability $[U(0.2,1)]$ can be obtained from the shaded areas of the PDFs shown in Fig.~\ref{fig:pdf_cp} above the values of the thresholds (black dashed vertical lines) in the corresponding plots.

{\it Viable regions in the network}: We define the viable region, $V_S$, of a source node $S$ for a given QIP task as the connected subgraph centered at $S$ within which the task can be performed with a viable probability of at least $\epsilon$. For practical entanglement-distribution protocols, the end-to-end connection quality are expected to decrease with increasing source-destination separation and therefore the viable probability, $\eta_l$, is a non-increasing function of the graph distance $l$. The viable region is defined as
\begin{align}
V_S&:=\{G_S\subseteq G ~|~ d_G(S,D)\leq r_*,~\forall D\in G_S\},\nonumber\\
r_*&:= \max\left\{l\,\middle|\, \eta_l\geq \epsilon \right\},
\label{viable_region}
\end{align}
where $r_*$ denotes the \textit{viable radius}. Thus, the viable region characterizes the portion of the network surrounding $S$ over which the given QIP task remains viable with confidence at least $\epsilon$. Different choices of the confidence level $\epsilon$ consequently define different viable regions and their corresponding viable radii.

A network region, $\tilde{V}_S$, around $S$ can be found where the path-parameters are satisfied with $100\%$ confidence and we call this region as the {\it absolute viable region} (AVR) of the source node, $S$, for a given QIP task. The AVR can be obtained by making the viability condition stringent via setting the value of $\epsilon$ to $1$ in Eq.~(\ref{viable_region}).% All the nodes within AVR is guaranteed to satisfy the multiparameter threshold for a QIP task with the source node. 
The radius of the AVR can be obtained in simple manner as,
\begin{align}
    \tilde{r}_*:=\max\left\{l\,\middle|\,\mu_{i,\min}^{(l)}\geq\mu_i^*\quad \forall i\in\{1,\ldots,n\}\right\}.
    \label{radius_avr}
\end{align}
Therefore, the radius of AVR does not depend of the nature of the PDFs of edge-parameters, rather they only depend on the minimum values of them. Therefore, the radius of AVR is a global source independent quantity. Moreover, two different edge-parameter distributions with the same minimum would yield the same radius for AVR. 
%\begin{comment}
\begin{figure}
    \centering
    \includegraphics[width=0.75\linewidth]{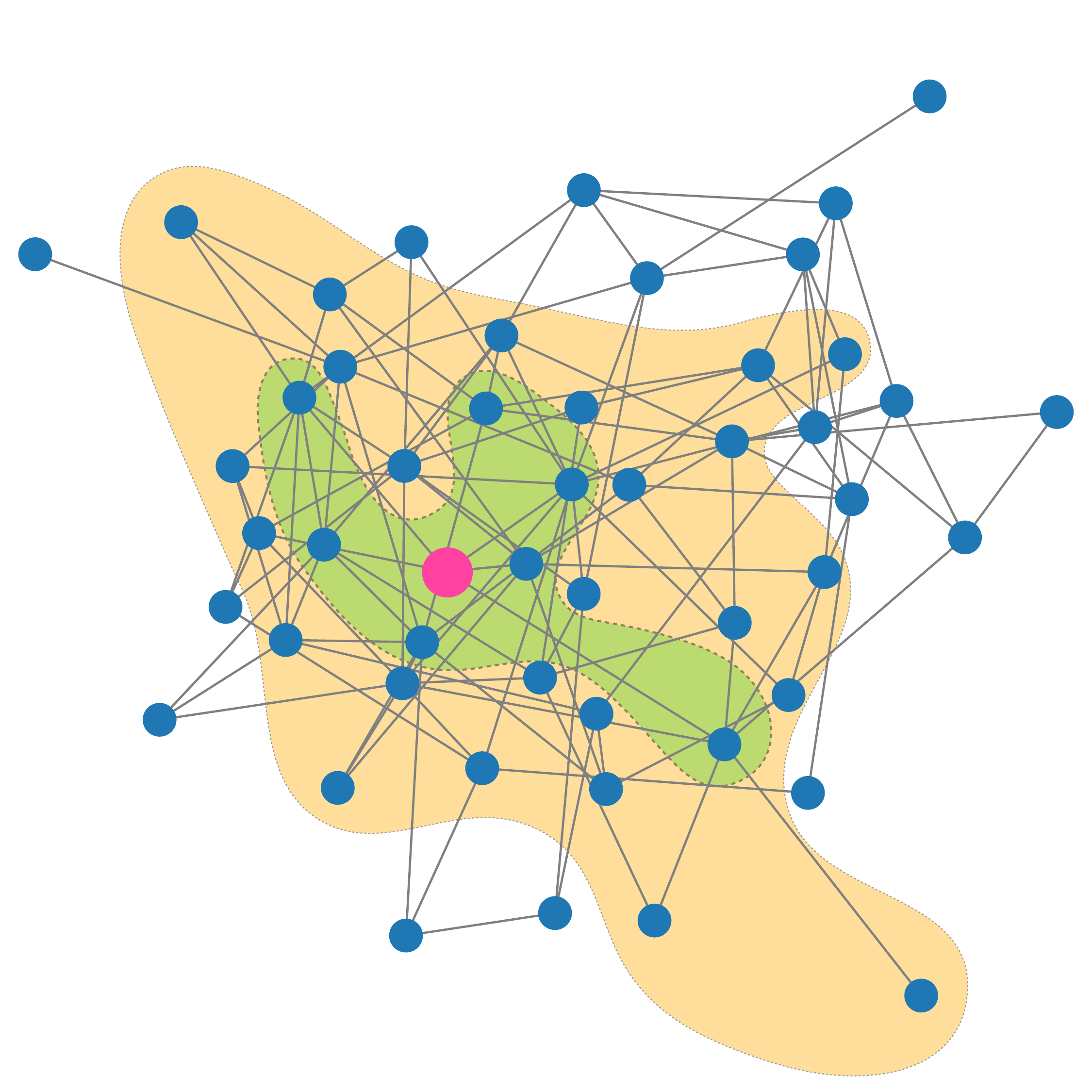}
    \caption{(color online) Schematic illustration of the AVR and PVR around a source node (magenta) in a complex quantum network. The AVR and PVR, shown as green and yellow shaded regions, respectively, have radii $\tilde{r}_*=1$ and $r_*=2$ for a given QIP task.}
    \label{fig:viab_regions_rn}
\end{figure}
%\end{comment}

Beyond the Absolute Viable Region (AVR), we identify a broader portion of the network where the multiparameter thresholds associated with the given QIP task are satisfied with a non-unit probability. We refer to this extended region as the {\it probabilistic viable region} (PVR). Note that, AVR is the smallest viable region situated at the core of different possible PVRs for different set of confidences, $\epsilon$; see Fig.~\ref{fig:viab_regions_rn} where the AVR and PVR corresponding to an arbitrary $\epsilon$ in a complex quantum network is schematically shown. Unlike the radius of AVR, the radius of the PVR can be different for different edge-parameter distributions in a given quantum network topology. Additionally, the radius of the PVR represents average over all possible source nodes in the network. Consequently, the radius is independent of any particular choice of source node and serve as global characteristics of the network for a given QIP task by providing an estimate of the graph distance up to which a source node can perform the task on average with a given minimum confidence.

{\it Viable node fraction in the network}: The viable probability has different values at different graph distances from the source node, $S$. Therefore, the probability, $\eta_l$, of a node satisfying the multiparameter threshold for the QIP task is dependent on the graph distance of the node from the source, $S$. The expected number of viable nodes within graph distance $d$ from the source node $S$, satisfying the multiparameter threshold for the QIP task can therefore be obtained as,
\begin{align}
\mathcal{N}_S(d)=\sum_{l=1}^{d} \eta_lN_{S|l},
\label{num_viable_nodes}
\end{align}
where $N_{S|l}$ denotes the total number of nodes at graph distance $l$ from the source node $S$. Note that $\mathcal{N}_S(d)$ in Eq.~(\ref{num_viable_nodes}) evaluates the expected number of viable nodes within any given graph distance $d$ as there is a non-zero probability of a node satisfying the multiparamter threshold even outside the PVR. However, PVR provides the network region where a node is capable of performing the QIP task with a guaranteed minimum confidence set by $\epsilon$. So, the expected proportion of viable nodes in the network for a given source node, $S$ can be obtained as,
\begin{align}
\mathcal{F}_S:=\frac{\mathcal{N}_S(d_{\rm net})}{N-1}.
\label{VNF}
\end{align}
Where, $d_{\rm net}$ denotes the diameter of the network which is defined as the maximum possible distance between any two nodes in the network and $N$ is the total number of nodes within the network. We call the expected proportion of viable nodes in the network defined in Eq.~(\ref{VNF}) as the \textit{viable node fraction} (VNF) of the source node $S$. Note that, the VNF is not a global characteristic of the network, rather it depends on the particular choice of the source node as the neighborhood of different nodes are not similar in general in an arbitrary network topologies.

\textit{Specific quantum network model}:
The statistical method presented in the preceding discussion to charactarize the functionality of a quantum network in terms of AVR, PVR and VNF can be applied to any general network topology with an arbitrary edge-parameter distribution and for an arbitrary entanglement distribution protocol. However, to demonstrate the idea we consider a specific scenario where the edges are distributed with quantum states of the form,
\begin{align}
\rho_e(\nu_e)=(1-\nu_e)\ket{\phi^+}\bra{\phi^+}+\nu_e\frac{\mathbf{1_4}}{4},
\label{edge_state}
\end{align}
with $\ket{\phi^+}=(\ket{00}+\ket{11})/\sqrt{2}$, which are a one-parameter family of mixed states with $0\leq \nu_e\leq 1$, widely considered in several recent QN models \cite{PENstates, Santra_20192, QN_cap_trans,  QN_optimal_routing}. The concurrence, a bipartite-entanglement measure, of states in Eq.~(\ref{edge_state}) is linearly related to the parameter, $\nu_e$, via $c(\rho_e)=\text{max}(0,1-3/2\nu_e)$, therefore, such states are quantum entangled whenever, $\nu_e<2/3$, equivalently, $\bra{\phi^+}\rho_e\ket{\phi^+}>1/2$.

\begin{comment}
\begin{figure*}[t]
    \centering
    \includegraphics[width=\linewidth]{PRL submission/viab_rn_copy.pdf}
    \caption{(color online) (A) The viable probability for the QKD is plotted against the graph distance $l$ when edge-concurrence and edge-probability has PDFs $q_C^{(1)}=U(0.9,1)$ and $q_P^{(1)}=U(0.2,1)$ respectively. The vertical black dashed (dotted) lines denote the radius of the AVR (PVR) for $\epsilon=0.25$. (B) The viable node fraction, $\mathcal{F}_S$, in a random network is plotted against average edge-concurrence (keeping average edge-probability fixed at $\overline{p}=1$) and (C) against average edge-probability (keeping average edge-concurrence fixed at $\overline{c}=1$).}
    \label{fig:viab_rn}
\end{figure*}
\end{comment}

Moreover, we characterize the network edges by two parameters -- concurrence $(C)$ and probability $(P)$ of entanglement generation and assume that the two edge-parameters are independent random variables. Therefore, the PDFs of edge-concurrence and edge-probability can be considered independent and we take it to be a uniform distribution, $q_X^{(1)}=U[x_\text{min},x_\text{max}]$ for $X\in\{C,P\}$ respectively. We also take a simple entanglement distribution protocol which is executed via performing entanglement swapping along the shortest graph path between any two nodes; for this specific protocol and statistically homogeneous edge-parameters, SGP provides the statistically optimal single-path connection as well. Moreover, the considered protocol preserves the form of the end-to-end state as Eq.~(\ref{edge_state}). %As the considered protocol uses a single path between two nodes for entanglement distribution, we denote the path-parameters of the end-to-end entangled states as path-parameters from now on.

\begin{figure}
    \centering
    \includegraphics[width=\linewidth]{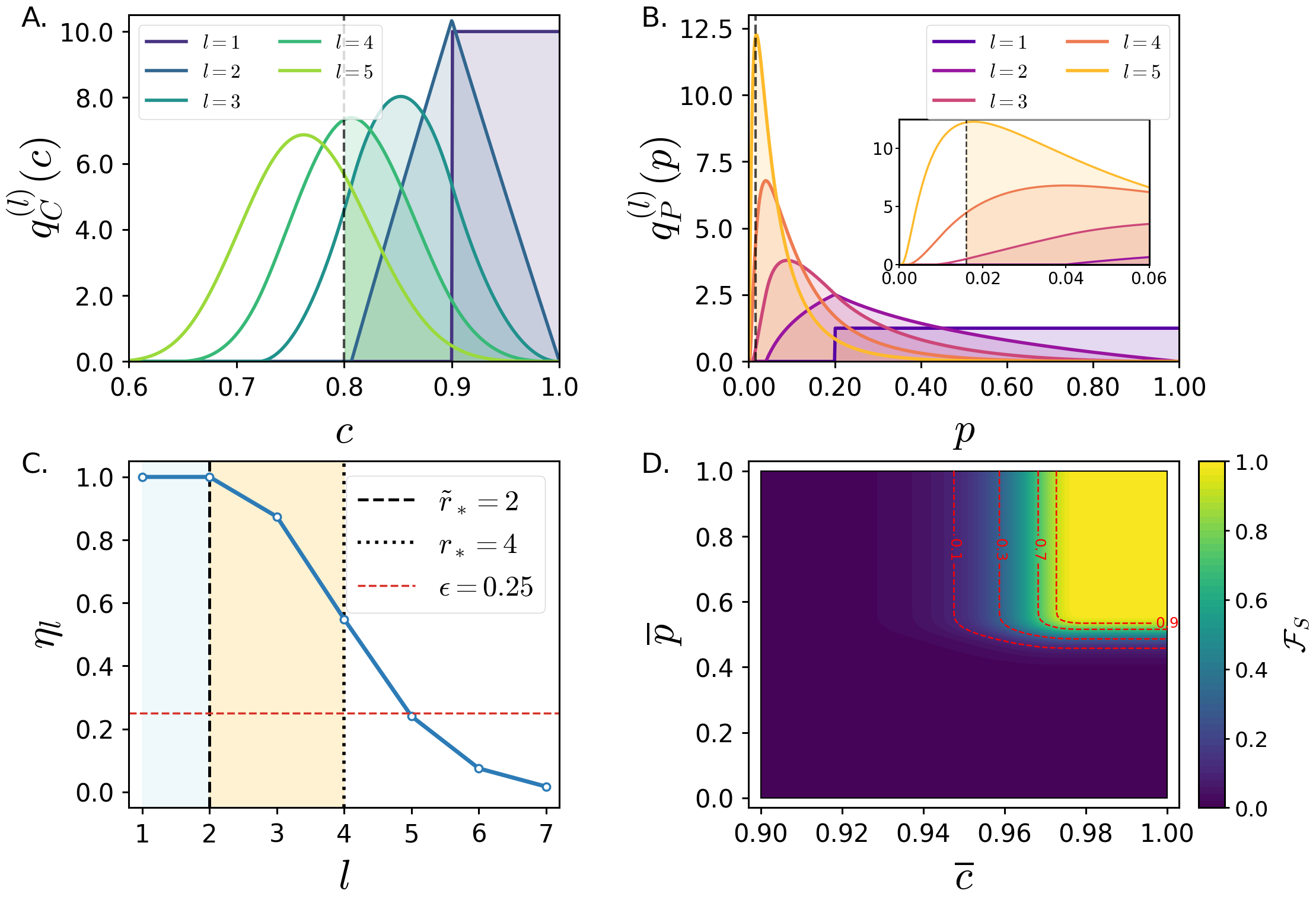}
    \caption{(color online) (A) PDFs of end-to-end concurrence $C^{(l)}$ for $l\in\{1,2,3,4,5\}$ with edge-concurrence PDF $q_C^{(1)}= U(0.9,1)$; shaded areas show the probability of $C^{(l)}\geq c_*=0.8$. (B) Corresponding PDFs of end-to-end probability $P^{(l)}$ with edge-probability PDF $q_P^{(1)}=U(0.2,1)$; shaded areas show the probability of $P^{(l)}\geq p_*=0.016$. (C) The viable probability $\eta_l$ for QKD plotted against the graph distance $l$ when edge-concurrence and edge-probability follow the uniform PDFs $q_C^{(1)}=U(0.9,1)$ and $q_P^{(1)}=U(0.2,1)$, respectively. The vertical black dashed (dotted) line denotes the radius of the AVR (PVR) for confidence $\epsilon=0.25$. (D) Heatmap of the viable node fraction, $\mathcal{F}_S$, in an Erd\H{o}s--R\'enyi random network as a function of the average edge-concurrence $\overline{c}$ and average edge-probability $\overline{p}$ with edge PDFs $q_X^{(1)}=U(x_{\min},x_{\max})$ and means $\overline{X}=(x_{\min}+x_{\max})/2$ for $X \in \{C, P\}$. The mean values are varied while keeping the support width $(x_{\max}-x_{\min})$ fixed at $0.2$, with symmetric boundary adjustments near $\overline{x}=0,1$ to constrain parameter values within $[0,1]$. Red contour lines denote iso-viability levels of $\mathcal{F}_S$ from $0.1$ to $0.9$.}
    \label{fig:viab_rn}
\end{figure}

In such a QN, the path-parameters achieved by ES at all intermediate nodes along the SGP of length $l$ is given by \cite{swapping_sohel},
\begin{align}
    C^{(l)}&=\max\left\{0,\frac{3}{2}\prod_{k=1}^{l}\frac{1+2C_k}{3}-\frac{1}{2}\right\},
    P^{(l)}=\prod_{k=1}^{l}P_k,
    \label{pathparameters}
\end{align}
where, $C_k$ and $P_k$ denote the concurrence and the probability of the $k^\text{th}$ edge along the path. In this case the PDF of the path-parameters can be evaluated from the PDF of the edge-parameters using Fourier convolution theorem \cite{arfken2011mathematical,mpep_mondal}. In the supplemental material, we describe a general prescription for deriving the PDF of the path-parameter, $X^{(l)}$, corresponding to an edge-parameter $X$. The scaling of the PDFs, $q_X^{(l)}$, of the path-parameters for different lengths of SGPs in this case is shown in Fig.~\ref{fig:viab_rn}\textcolor{red}{A},\textcolor{red}{B} and the exact analytical expressions are derived in the supplemental material. It is important to note that the viable probability monotonically decreases with the path length.

We consider QKD as an example of the QIP task that the node pairs want to perform in the network. For QKD a concurrence of, $C_{SD}\geq 0.78$, between a {\it S-D} pair is needed to obtain a non-zero finite secure key rate \cite{Santra_20192}. Assuming that the entangled photon sources in the network  operate at a rate of, $R_{eps}=1$MHZ, the network edge states are isotropic with concurrence, $C=0.80$, and the desired rate of secure key generation is, $R_{sec}=1$kHZ, then the probability of the entanglement connection needs to be at least, $P\geq R_{sec}/\{R_{eps}*(1-2h[(1-C)/2)]\}\simeq 1.6\%$, where $h[x]$ is the binary entropy function. Thus, for QKD entanglement connections with, $(C^{(l)},P^{(l)})\geq (c_*,p_*)=(0.8,1.6\%)$, between any {\it S-D} pair will be necessary and such {\it D}-nodes which satisfy the desired thresholds are part of the viable region for QKD relative to a given {\it S}.

\textit{Viability analysis in the considered network}: Due to the independence assumption of edge-concurrence and edge-probability, the viable probability can be written as, $\eta_l=\int_{c^*}^{c_\text{max}^{(l)}}q_C^{(l)}dc\int_{p^*}^{p_\text{max}^{(l)}}q_P^{(l)}dp$. Further the radius of the AVR in this case can be calculated exactly using Eq.~(\ref{radius_avr}) as,
\begin{align}
    \tilde{r}_*=\min\left\{ \big\lfloor\ln\left(\frac{1+2c^*}{3}\right)\middle /\ln\left(\frac{1+2c_\text{min}}{3}\right)\big\rfloor, \big\lfloor \frac{\ln p^*}{\ln p_\text{min}}\big\rfloor\right\}.
\end{align}
Therefore the radius of the AVR is determined solely by the QIP task threshold $(c^*,p^*)$ and the minimum of the edge-concurrence and edge-probability. For QKD, incorporating the values of the thresholds in the above equation, we obtain the radius of the AVR to be $\tilde{r}_*=2$ which is also shown in Fig.~\ref{fig:viab_rn}\textcolor{red}{C}. However, the radius of the PVR needs to be evaluated from the exact PDFs of the path-parameters using Eq.~(\ref{viable_region}). In Fig.~\ref{fig:viab_rn}\textcolor{red}{C} we have plotted the viable probability at different graph distances for QKD. Setting the confidence, $\epsilon=0.25$, we obtain the radius of the PVR as $r_*=4$. Therefore, the radius of AVR and PVR in this specific network model does not depend on the underlying topology; this in general is not true for an arbitrary network model with an arbitrary entanglement distribution protocol. However, the VNF is always dependent on the underlying network topology. Next we consider Erd\H{o}s--R\'enyi network topology and describe the scaling of VNF around an arbitrary source node.

{\it Scaling of viable node fraction in Erd\H{o}s--R\'enyi topology}:
Consider the above quantum network model based on an underlying Erd\H{o}s--R\'enyi random graph topology, $G(V,E)$, where $V$ is the set of nodes and $E$ is the set of edges. The number of nodes in the network is, $N=|V|=10^6$ and number of edges is, $|E|=5\times10^6$. The average degree of the nodes in the network can be calculated as, $k=2|E|/N=10$. In a random network each node can be treated identically as the degree of the nodes is close to the average degree of the network. Therefore the number of nodes at graph distance, $l$, around any source node, $S$, in the random network can be approximately estimated as, $N_{S|l}\approx k^l$ for $l\leq d_{\rm net}$, where $d_{\rm net}\sim\ln N/\ln k$ is the diameter of the network \cite{Barabasi@network}.

Applying our viability analysis to an Erd\H{o}s--R\'enyi random network topology we show in Fig.~\ref{fig:viab_rn}\textcolor{red}{D} the heatmap of the viable node fraction, $\mathcal{F}_S$, across the parameter space of average edge-concurrence $\overline{c}$ and average edge-probability $\overline{p}$. Over a significant regime (purple) of $(\overline{c}, \overline{p})$, the VNF remains practically vanishing ($\mathcal{F}_S \approx 0$), before crossing a sharp threshold boundary beyond which it rapidly saturates to unity (yellow region). 

Note that this sharp increase in VNF does not represent a thermodynamic phase transition, but rather a finite-size sharp threshold arising from the network topology. In a random network with average degree $k=10$, the vast majority of nodes, a fraction $(k-1)/k \sim 90\%$, reside at the boundary of the network. Consequently, the viable node fraction $\mathcal{F}_S = \sum_{l=1}^{d_{\text{net}}} \eta_l N_{S|l} / (N-1)$ is overwhelmingly dominated by the viable probability at the boundary, $\eta_{d_{\text{net}}}$. Because the viable probability decays with path length $l$, $\eta_{d_{\text{net}}}$ remains virtually zero across a large range of $(\overline{c}, \overline{p})$, suppressing $\mathcal{F}_S \approx 0$. As soon as the edge-parameters improve past the threshold where $\eta_{d_{\text{net}}}$ becomes non-zero and increases, $\mathcal{F}_S$ immediately follows and saturates toward unity. In the large network limit ($N \to \infty$), the diameter diverges, shifting both critical threshold values toward unity ($\overline{c}_c \to 1, \overline{p}_c \to 1$). Therefore, the observed jump is a topological cutoff effect governed by $\eta_{d_{\text{net}}}$ rather than a thermodynamic phase transition.

{\it Discussion and conclusions}: We have developed a statistical method to characterize the functional size of a quantum network for a given quantum information processing (QIP) task. The absolute and probabilistic viable regions quantify the network range over which the task can be performed with certainty and with a specified confidence, respectively, while the viable node fraction (VNF) gives the proportion of network nodes that can perform the task with a given source node. Together, these quantities provide a statistical characterization of the functionality of large-scale quantum networks based on the requirements of a QIP task.

An immediate extension of our method is to quantum repeater networks \cite{azuma2023quantum}, where different entanglement distribution protocols can be considered. For a given distribution of the initial edge-parameters, one can determine the protocol that provides the highest VNF, thereby identifying the distribution scheme that gives the largest functional size of the network. This provides a way to compare and optimize different entanglement distribution protocols at the network level.

The viable regions provide a complementary and more task-oriented characterization of quantum networks beyond entanglement percolation \cite{Acin2007}. While percolation provides a theoretically rich framework for identifying the edge-parameter regimes required to establish connectivity across opposite boundaries of a large network, practical QIP tasks may only require a local or smaller functional region around a source node. The AVR and PVR directly characterize such regions, making them useful for assessing the practical functionality of a quantum network for a given QIP task.

The statistical framework can also be extended to distributed quantum sensing \cite{Zang2026}, where the quality of entanglement between sensor pairs can vary due to stochastic entanglement generation. It can then identify the sensor nodes with which a given source sensor can perform a specified sensing task. Similarly, the framework can be used for task-specific network design and resource allocation by identifying the network regions and links required to support a given QIP task.

{\it Acknowledgements}: SS acknowledges useful conversations with David Elkouss at OIST and funding from DST, Govt. of India through the SERB grant MTR/2022/000389, IITB TRUST Labs grant DO/2023-SBST002-007 and the IITB seed funding.

\bibliographystyle{apsrev4-1}
\bibliography{refs-qnet}

%\newpage
\section*{Supplemental Material}
\section{Path-parameters of an entanglement connection}

The entanglement connection along any network path is realized via entanglement swapping at all intermediate stations. In our case two path parameters are of interest: the path concurrence and the path probability, both of which we derive in the following.\\
{\it Path concurrence}: At a particular node swapping can be performed on states, $\rho_{e_1}, \rho_{e_2} $, both of the form,
\begin{align}
\rho(\nu_{e_{1,2}})=(1-\nu_{e_{1,2}})\ket{\phi^+}\bra{\phi^+}+\nu_{e_{1,2}}\frac{\mathbf{1_4}}{4},
\label{supp_edge_states}
\end{align}
by projectively measuring the two qubits at the common node between the edges, one from each entangled pair, in a complete basis of the Bell states $\ket{\beta}=\ket{\phi_{\pm}},\ket{\psi_{\pm}}$ where $\ket{\phi_{\pm}}=(\ket{00}\pm\ket{11})/\sqrt{2}$ and  $\ket{\psi_{\pm}}=(\ket{01}\pm\ket{10})/\sqrt{2}$. This results in the post swapped state $\rho(\tilde{\nu})$ of the same form (up to local unitaries) as Eq.~(\ref{supp_edge_states}) but with the effective state parameter,
\begin{align}
 \tilde{\nu}=(\nu_{e_1}+\nu_{e_2}-\nu_{e_1}\nu_{e_2})=1-(1-\nu_{e_1})(1-\nu_{e_2}),
 \label{supp_q_two_edges}
 \end{align}
 All four outcomes corresponding to the different $\ket{\beta}$ are equiprobable and all four outcomes have the same concurrence of the post-swap state given by $C(\rho_{swapped})=(1-\frac{3}{2}\tilde{\nu})$. This calculation is done explicitly in reference \cite{entswap_conc}.

For evaluating the end-to-end concurrence of a path, $\mathcal{P}$, comprised of $l$-edges obtained by entanglement-swapping at all intermediate stations, we iterate Eq. (\ref{supp_q_two_edges}) to obtain the effective state parameter for path of length $l$ with states $\rho_{e_i},i=1,...,l$ along the edges as,
\begin{align}
\nu_{\mathcal{P}}=1-\prod_{i=1}^l(1-\nu_{e_i}).
\end{align}
Therefore concurrence of the path $\mathcal{P}$ can be obtained as,
\begin{align}
    C^{(l)}&=\text{Max}(0, 1-\frac{3}{2}\nu_\mathcal{P})\nonumber\\
    &=\text{Max}\left(0,1-\frac{3}{2}\bigg[1-\prod_{i=1}^l(1-\nu_{e_i})\bigg]\right)\nonumber\\
    &=\text{Max}\left(0,1-\frac{3}{2}\bigg[1-\prod_{i=1}^l(1-\frac{2}{3}(1-C_i))\bigg]\right)\nonumber\\
    &=\text{Max}\left(0, \frac{3}{2}\prod_{i=1}^{l}\frac{1+2C_i}{3}-\frac{1}{2}\right),
\end{align}
where, the concurrence of the state $\rho_i$ is $C_i$, i.e., $C_i=C(\rho_{e_i})$.

\begin{figure}
\centering
\includegraphics[width=\columnwidth]{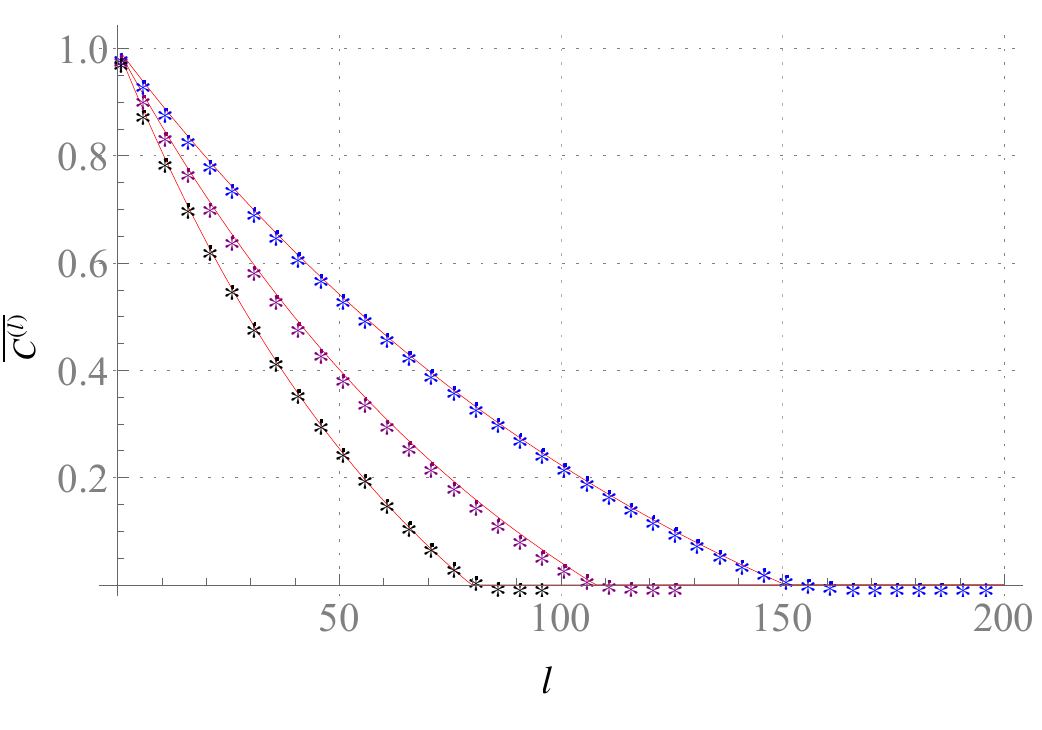}
%\vspace{-1cm}
\caption{(left) Simulated average concurrence (stars) vs path length in networks with different entanglement radii: $r^C=164, \overline{C}=0.99$ (Blue), $r^C=109, \overline{C}=0.985$ (Purple), $r^C=82, \overline{C}=0.98$ (Black). Smooth red line shows the analytical result. In all cases $10^4$ random paths were sampled.}
\label{supp_fig:aveconc}
\end{figure}

When the edge-concurrence takes values from a narrow distribution, we can make the approximation, $C_i\approx \overline{C}, \forall i$, where $\overline{C}$ is the average edge-concurrence obtained from the distribution. Therefore, the average path-concurrence in the network can be obtained as,
\begin{align}
\overline{C^{(l)}}=\text{Max}[0,\frac{3}{2}\frac{(1+2\overline{C})^l}{3^l}-\frac{1}{2}].
\end{align}
Considering the parametrization, $\overline{C}=(1-\delta_1)$, we can also write,
\begin{align}
\overline{C^{(l)}}=\text{Max}[0,\frac{3}{2}(1-(2/3)\delta_1)^l-\frac{1}{2}].
\end{align}
For large-scale networks of size $N\to\infty$, we need $\delta_1\sim 1/\text{Poly}(N)\to 0$. Therefore, we get the behavior of $\overline{C^{(l)}}$ for short- and long- path lengths, $l$ as,
\begin{align}
\overline{C^{(l)}}&\simeq (1-l\delta_1),~~ l\ll(1/\delta_1)=\text{Poly}(N)\nonumber\\
\overline{C^{(l)}}&= \frac{3}{2}e^{-(2/3)\delta_1l}-\frac{1}{2},~~l\lesssim\frac{3}{2}\frac{\ln 3}{\delta_1}:=r^C\sim \text{Poly}(N)
\label{supp_path_conc}
\end{align}
Notice that the average concurrence for a path decreases monotonically with its length. The decrease is linear for short graph distances but exponentially suppressed at path lengths that are comparable to the entanglement radius of the network. This is shown in Fig. \ref{supp_fig:aveconc}.

{\it Path concurrence}: The path probability in the entanglement distribution scheme is simply a product of the edge probabilities $P_{\mathcal{P}}=\prod_e p_e$, therefore the path-probability for a path of length $l$ where the edge probabilities are i.i.d. is,
\begin{align}
P^{(l)}=\prod_{i=1}^l p_{e_i}.
\end{align}
Considering an approximation for the average probability $P_i\approx \overline{P},\forall i$ in the case of narrow distribution of the edge-probability and the parameterization $\overline{P}=(1-\delta_2)$ similar to concurrence, we obtain the average path-probability as,
\begin{align}
    \overline{P^{(l)}}=(1-\delta_2)^l.
\end{align}
The behavior of $\overline{P^{(l)}}$ for short- and long- path lengths, $l$, is thus,
\begin{align}
\overline{P^{(l)}}&\simeq (1-l\delta_2),~~ l\ll(1/\delta_2)=\text{Poly}(N)\nonumber\\
\overline{P^{(l)}}&\simeq e^{-\delta_2l},~~ l\simeq(1/\delta_2)=\text{Poly}(N)\nonumber\\
\label{supp_path_prob}
\end{align}
Again, notice that the average probability for a path decreases monotonically with its length. The decrease is linear for short graph distances but exponentially suppressed at path lengths that are comparable to the connection radius of the network.

%%%%%%%%%%%%%%%%%%%%%%%%%%%%%%%%%%%%%%%

%%%%%%%%%%%%%%%%%%%%%%%%%%%%%%%%
\section{PDF of the path-parameters} \label{pdf_path_appen}
Here we provide the detailed analytical prescription for deriving the probability distribution function (PDF), $q_X^{(l)}$, of the path-parameters, $X^{(l)}$, from the PDF, $q_X^{(1)}$, of the microscopic edge-parameters, $X$. We consider an entanglement distribution scheme where the end-to-end parameter across a path of length $l$ combines multiplicatively under a suitable change of variables. 

\subsection{General transform and convolution framework}
Define a random variable $W = f(X)$ such that the effective path-parameter across an $l$-hop path satisfies the product relation:
\begin{align}
    W^{(l)} = \prod_{k=1}^l W_k,
\end{align}
where $W_k = f(X_k)$ represents the transformed random variable associated with the $k^\text{th}$ edge along the path. The PDF of the single-edge variable $W$ is obtained via the Jacobian transformation as $q_W(w) = q_X(f^{-1}(w)) \left|\frac{d}{dw}f^{-1}(w)\right|$. 

To convert the product of random variables into an additive sum, we apply a logarithmic mapping $Y = \ln W$ and define the path sum variable:
\begin{align}
    T = \ln W^{(l)} = \sum_{k=1}^l Y_k, \quad \text{where } Y_k = \ln W_k.
\end{align}
The PDF of $Y$ is given by $q_Y(y) = e^y q_W(e^y)$. For statistically independent and identically distributed (i.i.d.) network edges, the PDF of the sum $T$ is given by the $l$-fold convolution of $q_Y(y)$:
\begin{align}
    q_T(t) = (q_Y * q_Y * \dots * q_Y)(t),
\end{align}
where $*$ denotes the continuous convolution operator, $(f * g)(t) := \int_{-\infty}^{\infty} f(\tau) g(t - \tau) d\tau$. By the convolution theorem, the Fourier transform of an $l$-fold convolution equals the $l^\text{th}$ power of the individual Fourier transform:
\begin{align}
    \mathcal{F}\{q_T(t)\} = \left[ \mathcal{F}\{q_Y(y)\} \right]^l.
\end{align}
The PDF $q_T(t)$ is subsequently obtained by evaluating the inverse Fourier transform:
\begin{align}
    q_T(t) = \frac{1}{2\pi} \int_{-\infty}^{\infty} e^{-it\omega} \left[ \int_{-\infty}^\infty e^{i\omega y} q_Y(y) dy \right]^l d\omega.
    \label{eq:inv_FT_general}
\end{align}
Once $q_T(t)$ is determined, the exact PDF of the physical path-parameter $X^{(l)}$ is recovered by inverting the transformation:
\begin{align}
    q_X^{(l)}(x) = \frac{1}{f(x)} q_T(\ln f(x)) \left| \frac{d}{dx}f(x) \right|.
    \label{eq:pdf_path_param_formal}
\end{align}

\subsection{Analytical derivation for uniformly distributed edge-parameters}
We consider the general case where the edge-parameter $X$ is distributed uniformly over the bounded interval $[x_{\rm min}, x_{\rm max}]$ according to $q_X^{(1)}(x) = \frac{1}{x_{\rm max} - x_{\rm min}}$. Under the linear mapping $W = f(X)$, the transformed variable $W$ is uniformly distributed over $[a, b]$, where $a = f(x_{\rm min})$ and $b = f(x_{\rm max})$, with density $q_W(w) = \frac{1}{b - a}$.

%\subsubsection{Characteristic function of a single link}
The single-edge log-variable $Y = \ln W$ is supported on $[\ln a, \ln b]$ with PDF $q_Y(y) = \frac{e^y}{b - a}$. Its Fourier transform (characteristic function) is evaluated directly as:
\begin{align}
    \phi_Y(\omega) &= \int_{\ln a}^{\ln b} e^{i\omega y} \left(\frac{e^y}{b - a}\right) dy \nonumber\\
    &= \frac{1}{b - a} \int_{\ln a}^{\ln b} e^{(1 + i\omega)y} dy = \frac{b^{1 + i\omega} - a^{1 + i\omega}}{(1 + i\omega)(b - a)}.
\end{align}
For an $l$-edge path, the characteristic function of the sum $T$ becomes:
\begin{align}
    \phi_T(\omega) = [\phi_Y(\omega)]^l = \frac{1}{(b - a)^l} \frac{\left( b^{1 + i\omega} - a^{1 + i\omega} \right)^l}{(1 + i\omega)^l}.
    \label{eq:phi_T_expr}
\end{align}
Applying the binomial theorem to the numerator of Eq.~(\ref{eq:phi_T_expr}) yields:
\begin{align}
    \left( b^{1 + i\omega} - a^{1 + i\omega} \right)^l &= \sum_{n=0}^l (-1)^n \binom{l}{n} (b^{1+i\omega})^{l-n} (a^{1+i\omega})^n \nonumber\\
    &= \sum_{n=0}^l (-1)^n \binom{l}{n} \nu_n e^{i\omega \tau_n},
\end{align}
where we have defined the boundary parameters:
\begin{align}
    \nu_n \equiv b^{l-n} a^n, \quad \tau_n \equiv \ln \nu_n = (l-n)\ln b + n \ln a, \nonumber\\
    \quad n \in \{0, 1, \dots, l\}.
    \label{eq:boundary_definitions}
\end{align}

%\subsubsection{Inverse Fourier transform via contour integration}
Substituting the expanded characteristic function into the inverse Fourier transform Eq.~(\ref{eq:inv_FT_general}) gives:
\begin{align}
    q_T(t) = \frac{1}{(b - a)^l} \sum_{n=0}^l (-1)^n \binom{l}{n} \nu_n \left[ \frac{1}{2\pi} \int_{-\infty}^{\infty} \frac{e^{i\omega(\tau_n - t)}}{(1 + i\omega)^l} d\omega \right].
    \label{eq:qT_integral_sum}
\end{align}
The integral within the brackets, $I_n(t) \equiv \frac{1}{2\pi} \int_{-\infty}^\infty \frac{e^{i\omega(\tau_n - t)}}{(1 + i\omega)^l} d\omega$, is evaluated using Cauchy's residue theorem in the complex $\omega$-plane. The integrand possesses an $l^\text{th}$-order pole at $\omega = i$:
\begin{itemize}
    \item For $t < \tau_n$ ($\tau_n - t > 0$), Jordan's lemma permits closing the integration contour in the upper half-plane ($\text{Im}(\omega) > 0$), which encloses the pole at $\omega = i$. The residue is:
    \begin{align}
        \text{Res}_{\omega = i} \left[ \frac{e^{i\omega(\tau_n - t)}}{i^l (\omega - i)^l} \right] &= \frac{1}{(l-1)!~ i^l}~ \lim_{\omega \to i} \frac{d^{l-1}}{d\omega^{l-1}} e^{i\omega(\tau_n - t)} \nonumber\\
        &= \frac{(\tau_n - t)^{l-1} e^{-(\tau_n - t)}}{i(l-1)!}.
    \end{align}
    Multiplying by $2\pi i$ and the prefactor $\frac{1}{2\pi}$ yields:
    \begin{align}
        I_n(t) = \frac{(\tau_n - t)^{l-1}}{(l-1)!} e^{-(\tau_n - t)}.
    \end{align}
    \item For $t > \tau_n$ ($\tau_n - t < 0$), the contour must be closed in the lower half-plane ($\text{Im}(\omega) < 0$). Since the lower half-plane contains no singularities, Cauchy's theorem gives $I_n(t) = 0$.
\end{itemize}
Thus, $I_n(t)$ is compactly expressed with the Heaviside step function $\Theta(\cdot)$:
\begin{align}
    I_n(t) = \frac{(\tau_n - t)^{l-1}}{(l-1)!} e^{-(\tau_n - t)} \Theta(\tau_n - t).
\end{align}
Noting that $\nu_n e^{-(\tau_n - t)} = e^{\tau_n} e^{-\tau_n + t} = e^t$, Eq.~(\ref{eq:qT_integral_sum}) simplifies to:
\begin{align}
    q_T(t) = \frac{e^t}{(l - 1)! (b - a)^l} \sum_{n=0}^l (-1)^n \binom{l}{n} (\tau_n - t)^{l-1} \Theta(\tau_n - t).
\end{align}
For any evaluation point $t$ located within the $m^\text{th}$ interval $[\tau_m, \tau_{m-1}]$, the step function satisfies $\Theta(\tau_n - t) = 1$ for $n \le m - 1$ and vanishes for $n \ge m$. Hence, the sum truncates at $m - 1$:
\begin{align}
    q_T(t) = \frac{e^t}{(l - 1)! (b - a)^l} \sum_{n=0}^{m-1} (-1)^n \binom{l}{n} \left[ \ln\left(\frac{b^{l-n} a^n}{e^t}\right) \right]^{l-1}.
\end{align}
Transforming back to the multiplicative path variable $w = e^t$ via $q_{W^{(l)}}(w) = \frac{1}{w} q_T(\ln w)$ gives:
\begin{align}
    q_{W^{(l)}}(w) = \frac{1}{(l - 1)! (b - a)^l} \sum_{n=0}^{m-1} (-1)^n \binom{l}{n} \left[ \ln\left(\frac{b^{l-n} a^n}{w}\right) \right]^{l-1},
    \label{eq:qW_piecewise}
\end{align}
Here, the upper summation limit $m-1$ depends on the interval in which the evaluation point $w$ lies. For $m\in\{1,2,\ldots,l\}$, if
\begin{align}
b^{\,l-m}a^m \leq w \leq b^{\,l-m+1}a^{m-1},
\end{align}
then the terms with $n\geq m$ vanish because of the Heaviside function, and the summation therefore runs from $n=0$ to $n=m-1$.

\subsection{Exact closed-form expressions for path-concurrence and path-probability}
In our quantum network model, the two relevant parameters are concurrence ($C$) and the probability of entanglement generation ($P$), whose end-to-end values along a shortest graph path of length $l$ are given by Eq.~(\textcolor{red}{10}) in the main text.

\subsubsection{Path probability distribution $q_P^{(l)}(p)$}
For link probabilities distributed uniformly as $P \sim U(p_{\rm min}^{(1)}, p_{\rm max}^{(1)})$, the transformation function is the identity $f(P) = P$. Setting $w = p$, $a = p_{\rm min}^{(1)}$, and $b = p_{\rm max}^{(1)}$ in Eq.~(\ref{eq:qW_piecewise}) directly yields:
\begin{align}
    q_P^{(l)}(p) &= \frac{1}{(l-1)!} \left(\frac{1}{p_{\rm max}^{(1)} - p_{\rm min}^{(1)}}\right)^l\nonumber\\ &\times\sum_{n=0}^{m-1} (-1)^n \binom{l}{n} \left[ \ln\left( \frac{(p_{\rm max}^{(1)})^{l-n} (p_{\rm min}^{(1)})^n}{p} \right) \right]^{l-1},
    \label{eq:prob_pdf_gen}
\end{align}
where the piecewise index $m \in \{1, 2, \dots, l\}$ is determined by the interval containing $p$:
\begin{align}
    p \in \left[ (p_{\rm max}^{(1)})^{l-m} (p_{\rm min}^{(1)})^m, \;\; (p_{\rm max}^{(1)})^{l-m+1} (p_{\rm min}^{(1)})^{m-1} \right].
\end{align}

\subsubsection{Path concurrence distribution $q_C^{(l)}(c)$}
For edge concurrence distributed as $C \sim U(c_{\rm min}^{(1)}, c_{\rm max}^{(1)})$, the required linear transformation is $W = f(C) = (1 + 2C)/3$. The boundary limits map to $a = (1 + 2c_{\rm min}^{(1)})/3$ and $b = (1 + 2c_{\rm max}^{(1)})/3$, giving the denominator difference:
\begin{align}
    b - a = \frac{2}{3}\left(c_{\rm max}^{(1)} - c_{\rm min}^{(1)}\right).
\end{align}
Applying the Jacobian scaling $\left| \frac{dw}{dc} \right| = \frac{2}{3}$, the prefactor becomes:
\begin{align}
    \left| \frac{dw}{dc} \right| \frac{1}{(b - a)^l} = \frac{2}{3} \frac{1}{\left[ \frac{2}{3}(c_{\rm max}^{(1)} - c_{\rm min}^{(1)}) \right]^l} = \left(\frac{3}{2}\right)^{l-1} \left(\frac{1}{c_{\rm max}^{(1)} - c_{\rm min}^{(1)}}\right)^l.
\end{align}
Substituting into Eq.~(\ref{eq:qW_piecewise}) via Eq.~(\ref{eq:pdf_path_param_formal}) yields:
\begin{align}
    q_C^{(l)}(c) &= \frac{1}{(l-1)!} \left(\frac{3}{2}\right)^{l-1} \left(\frac{1}{c_{\rm max}^{(1)} - c_{\rm min}^{(1)}}\right)^l\nonumber\\
    &\times\sum_{n=0}^{m-1} (-1)^n \binom{l}{n} \left[ \ln\left( \frac{(1+2c_{\rm max}^{(1)})^{l-n} (1+2c_{\rm min}^{(1)})^n}{(1+2c)\,3^{l-1}} \right) \right]^{l-1},
    \label{eq:conc_pdf_gen}
\end{align}
valid on the concurrence sub-intervals:
\begin{align}
    c \in& \left[ \frac{3}{2}\left(\frac{1+2c_{\mathrm{max}}^{(1)}}{3}\right)^{l-m}\left(\frac{1+2c_{\mathrm{min}}^{(1)}}{3}\right)^m - \frac{1}{2}, \right. \nonumber \\
    &\quad \left. \frac{3}{2}\left(\frac{1+2c_{\mathrm{max}}^{(1)}}{3}\right)^{l-m+1}\left(\frac{1+2c_{\mathrm{min}}^{(1)}}{3}\right)^{m-1} - \frac{1}{2} \right].
\end{align}
When the upper boundary is ideal ($x_{\rm max}^{(1)} = 1$), Eqs.~(\ref{eq:prob_pdf_gen}) and (\ref{eq:conc_pdf_gen}) reduce directly to the $U(x_{\rm min}, 1)$ formulas.

%%%%%%%%%%%%%%%%%%%%%%%%%%%%%%%%%%
\section{Radius of Absolute and Probabilistic Viable Regions}

\subsection{Absolute viable region} 
The radius of the absolute viable region is obtained by solving, $\eta_l=1$ for maximum $l$. Note that, the absolute viability condition is satisfied if and only if, $x_{\min}^{(l)}\geq x_*$ where $x_{\min}^{(l)}$ is the minimum value of the path-parameter for $x\in\{c,p\}$. The expressions for this minimum for path-concurrence and path-probability can be respectively given as,
\begin{align}
    c_{\min}^{(l)}&=\text{Max}\left[0,\frac{3}{2}\bigg(\frac{1+2c_{\min}}{3}\bigg)^l-\frac{1}{2}\right],
    ~p_{\min}^{(l)}=(p_{\min})^l.
\end{align}
Therefore the radii of the AVRs corresponding to path-concurrence and path-probability can be obtained by solving the two inequalities,
\begin{align}
    \frac{3}{2}\bigg(\frac{1+2c_{\min}}{3}\bigg)^l-\frac{1}{2}\geq c_*,~ p_{\min}^l\geq p_*.
\end{align}
Solving these two inequalities we obtain the respective radii as,
\begin{align}
    \tilde{r}_*=\min\left\{ \big\lfloor\ln\left(\frac{1+2c^*}{3}\right)\middle /\ln\left(\frac{1+2c_\text{min}}{3}\right)\big\rfloor, \big\lfloor \frac{\ln p^*}{\ln p_\text{min}}\big\rfloor\right\}.
\end{align}
where the floor values are taken as the graph distances can only take integer values. It is important to note from the above derivation that the radius of the AVR does not depend of the explicit form of the PDF of the edge-parameters, rather the minimum values of the edge-parameters being above the respective thresholds suffices for the existence of an AVR in a network.

\subsection{Probabilistic viable region} 
The evaluation of the radius of PVR requires the knowledge of the PDF of the edge-parameters and evaluated PDF of the path-parameters unlike the radius of the AVR. In our example of edge-parameters distributed according to uniform distributions, the radius of the PVR can be obtained by solving $\eta_l\geq\epsilon$ where $\eta_l$ can be evaluated by integrating the PDFs of the path-concurrence and path-probability obtained in Eqs.~(\ref{eq:prob_pdf_gen}) and (\ref{eq:conc_pdf_gen}). We have performed the integration numerically and plotted the results in the main text.
%%%%%%%%%%%%%%%%%%%%%%%%%%%%%%%%
\end{document}